\documentclass[preprint,amsmath,amssymb,aps]{revtex4-1}

\usepackage{graphicx}
\usepackage{dcolumn}
\usepackage{bm}
\usepackage{braket}

\usepackage{amsmath}	
\begin{document}

\title{Shape of $^{\bf 13}{\bf C}$ studied by the  real-time evolution method}
\author{S. Shin}
 \affiliation{Department of Physics, Hokkaido University, Sapporo 060-0810, Japan}
\author{B. Zhou}
 \affiliation{Institute of Modern Physics, Fudan University, Shanghai 200433, China}
\author{M. Kimura}
 \email{masaaki@nucl.sci.hokudai.ac.jp}
 \affiliation{Department of Physics, Hokkaido University, Sapporo 060-0810, Japan}
 \affiliation{Nuclear Reaction Data Centre, Hokkaido University,
  Sapporo 060-0810, Japan} 
  \affiliation{Research Center for Nuclear Physics (RCNP), Osaka University,
  Ibaraki 567-0047, Japan}

\date{\today}

\begin{abstract}
 \noindent$\bf{Background:}$ Recently, Bijker et al. [Phys. Rev. Lett. 122,  162501 (2019)]
 explained the rotation-vibration spectrum of $^{13}{\rm C}$ by assuming triangular nuclear shape
 with $D'_{3h}$ symmetry. \\  
 \noindent$\bf{Purpose:}$ The purpose of this work is to test the shape and symmetry of
 $^{13}{\rm C}$  based on  a microscopic nuclear model without assumption of nuclear  shape.\\  
 \noindent$\bf{Method:}$ We have applied the real-time  evolution method to  $^{13}{\rm C}$.
 By using the equation-of-motion of clusters, the model describes the 3$\alpha$+$n$ system without
 any assumption of symmetry.\\     
\noindent$\bf{Results:}$ REM described the low-lying states more accurately than the  previous
 cluster model studies. The analysis of the wave functions showed that the ground band has
 approximate triangular symmetry, while the excited bands deviate from it. \\   
\noindent$\bf{Conclusion:}$ This work confirmed that the ground band has the intrinsic structure
 with the triangular arrangement of three alpha particles. 
\end{abstract}

\pacs{Valid PACS appear here}
\maketitle


\section{Introduction}
The carbon isotopes have been an important subject in nuclear cluster physics as they manifest a
rich variety of cluster phenomena. The Hoyle state (the $0_2^+$ state of $^{12}$C) exhibits one of
the most interesting clustering aspects Bose-Einstein condensate (BEC) of three $\alpha$
particles~\cite{Tohsaki2001a}. The structure of the Hoyle state and its analogous states in
neighboring nuclei has been one  of the major topics in these decades
~\cite{Funaki2003,Wakasa2007,Chernykh2007,Kanada-Enyo2007,Funaki2008,Itoh2011,Epelbaum2011,
Epelbaum2012,Freer2012,Fukuoka2013,Ohtsubo2013,Carlson2015,Zhou2016}. In the highly
excited region of carbon isotopes, a different type of clustering, the linear-chain of alpha
particles, has also been intensively discussed
~\cite{Itagaki2001,Maruhn2010,Suhara2010,Baba2014,Ebran2014,Freer2014,Baba2016,Fritsch2016,
Baba2017,Yamaguchi2017,Li2017,Baba2018,Marevic2019,Liu2020}. Among the carbon isotopes,
$^{13}{\rm C}$ is of particular importance as the system composed of three $\alpha$ particles
(bosons) plus a valence nucleon (fermion). The Hoyle analog state in $^{13}{\rm C}$, the BEC of
three $\alpha$ particles with a neutron as an impurity, and the possible formation of the
linear-chain structure assisted by a valence neutron have been main interests for this
nucleus~\cite{Milin2002,Freer2011,Furutachi2011,Suhara2011,Wheldon2012,Yamada2015,Ebran2017,
Chiba2020,Inaba2020}.    

Recently, apart from these studies, Bijker et al.~\cite{Bijker2019} proposed a different
interpretation for the structure of $^{13}{\rm C}$ based on the symmetry arguments. They applied the
algebraic cluster model (ACM)~\cite{Bijker2000, DellaRocca2017, Bijker2020} to $^{13}{\rm C}$, and
assumed triangular $D'_{3h}$ symmetry of the $3\alpha$ clusters accompanied by a valence
neutron. The intrinsic states were classified into three representations of the symmetry group, and
each of them forms the rotational band and exhibits unique spectrum and  transition strengths. Based
on a comparison with available experimental data, they argued that many of the observed ground and
excited states can be assigned to these bands, and hence,  $^{13}{\rm C}$ has triangular $D'_{3h}$
symmetry. This suggests an interesting insight into the structure of carbon isotopes and contradicts 
the BEC interpretation of the Hoyle state and its analog states.  However, the model is based on
purely symmetry concepts and the deviation from the triangular symmetry, which must take place in
reality, is neglected. Therefore, the symmetry behind the spectrum of $^{13}{\rm C}$ and deviation
from it should be tested by the microscopic models without any assumption of the nuclear shape. 

The real-time evolution method (REM) recently proposed by Imai et al.~\cite{Imai2019} is one of the 
microscopic cluster models which can examine the shape of nuclei without any assumption of the
symmetry. It generates basis wave functions with various cluster configurations by using the
equation-of-motion (EOM) of the Gaussian wave packets. A benchmark calculation showed that REM
precisely describes the 3$\alpha$ system including the Hoyle state. Therefore, a natural idea is to
extend the method to the non-$N\alpha$ systems~\cite{Zhou2020}. It is noted that REM superposes massive
number of the basis wave functions to describe the cluster systems, and it does not introduce any
assumption about the symmetry of  nuclear shape and cluster configurations. Therefore, REM is a
suitable nuclear model to test if there exists any symmetry in the spectrum of
$^{13}{\rm C}$. Thus, the aim of this work is two-fold. The first is the extension and benchmark of
REM for non-$N\alpha$ system, and the second is the verification of the triangular $D'_{3h}$
symmetry in the  spectrum of $^{13}{\rm C}$.  

We organize this paper as follows. In the next section, the framework of REM for the 3$\alpha$+$n$
system is briefly explained. In the section III, the numerical results are presented. Compared to
a previous study which used the same Hamiltonian, REM yields deeper binding energies for all bound
states and describes the 3$\alpha$+$n$ system more accurately. Based on the $B(E2)$ strengths, we
propose an assignment of the rotational bands and discuss the internal structure of the band member
states to examine the triangular $D'_{3h}$ symmetry. It is shown that the ground band member states
have the same intrinsic structure which has triangular arrangement of three alpha
particles. However, it is found that many excited states deviate from a rigid body and fluctuate
around the triangular shape. Finally, in the last section, we summarize this work. 


\section{Theoretical framework}

In this section, we explain the Hamiltonian and  framework of the real-time evolution method for the
3$\alpha$+$n$ system. The Hamiltonian used in this study is given as, 
\begin{align}
 \hat{H} = \sum_{i=1}^{13} \hat{t}_i  + \sum_{i<j}^{13}\hat{v}_N(r_{ij}) +
 \sum_{i<j}^{13}\hat{v}_C(r_{ij}) - \hat{t}_{cm},  
 \label{eq:ham}
\end{align}
where $\hat{t}_i$ is the kinetic energy of the $i$th nucleon and $\hat{t}_{cm}$ is the
center-of-mass kinetic energy. The $\hat{v}_{N}$ and $\hat{v}_{C}$ denote the effective
nucleon-nucleon and Coulomb interactions, respectively. For the central part of the nucleon-nucleon
interaction, we used Volkov No.~2 force~\cite{Volkov1965} with the exchange parameters, $W=0.4$,
$B=H=0.125$ and $M=0.6$. The G3RS force~\cite{Yamaguchi1979} is used for the spin-orbit part with
two choice of the strengths $u_{ls}=1000$ and 2000 MeV. The latter value $u_{ls}=2000$ MeV was also
used by Furutachi et al.~\cite{Furutachi2011}, and we also adopt the same strength for the sake of 
comparison. However, as shown later, it does not reproduce the correct ordering of the ground band
member states, and hence, we also applied a weaker strength $u_{ls}=1000$ MeV for better description  
of the ground band.  
 
As the basis wave function to describe the $3\alpha+n$ system, we employ the Brink-Bloch wave
function~\cite{Brink1966} which consists of three $\alpha$ clusters with $(0s)^4$ configuration
coupled with a valence neutron, 
\begin{align}
 \Phi(\bm Z_1,...,\bm Z_4) &= \mathcal A
 \Set{\Phi_\alpha(\bm Z_1)\Phi_\alpha(\bm Z_2)\Phi_\alpha(\bm Z_3)\Phi_n(\bm Z_4)},
 \label{eq:brink1}\\ 
 \Phi_\alpha(\bm Z) &= \mathcal A
 \Set{\phi(\bm r_1,\bm Z)\chi_{p\uparrow}\cdots\phi(\bm r_4,\bm Z)\chi_{n\downarrow}},\\
 \Phi_n(\bm Z) &=\phi(\bm r,\bm Z)\chi_{n\uparrow},\label{eq:brink2}\\
 \phi(\bm r,\bm Z) &= \left(\frac{2\nu}{\pi}\right)^{3/4}\exp
 \set{-\nu\left(\bm r- \bm Z\right)^2},
\end{align}
where $\Phi_\alpha(\bm Z)$ and $\Phi_n(\bm Z)$ denote the wave packets describing the $\alpha$
cluster and the valence neutron located at $\bm Z$, respectively. In this study, we fix the valence
neutron spin to up in the intrinsic frame without loss of generality. The set of three-dimensional
vectors $\bm Z_1,...,\bm Z_4$ is complex numbered and describes positions and momenta of the
$3\alpha+n$ clusters in the phase space. The size parameter $\nu=1/2b^2$ of the $\alpha$ particle is
fixed to $b=$ 1.46 fm~ which reproduces the observed size of an $\alpha$ particle. The same size
parameter is also used to describe the valence neutron. 

In the REM framework, we use the equation-of-motion to generate the basis wave functions with
various configurations of clusters. From the time-dependent variational principle,  
\begin{align}
 \delta\int dt\frac{\langle\Phi(\bm Z_1,...,\bm Z_4)|i\hbar\;d/dt-\hat{H}|
 \Phi(\bm Z_1,...,\bm Z_4)\rangle}{\langle\Phi(\bm Z_1,...,\bm Z_4)|
 \Phi(\bm Z_1,...,\bm Z_4)\rangle}=0, 
\end{align}
one obtains the equation-of-motion (EOM) for the $3\alpha+n$ cluster centroids
$\bm Z_1,...,\bm Z_4$,
\begin{align}
  i\hbar&\sum_{j=1}^4\sum_{\sigma=x,y,z} C_{i\rho j\sigma}\frac{dZ_{j\sigma}}{dt} =
 \frac{\partial \mathcal H_{int}}{\partial Z_{i\rho}^*}, \label{eq:eom}\\ 
 \mathcal H_{int}&\equiv\frac{\langle\Phi(\bm Z_1,...,\bm Z_4)|\hat{H}|
 \Phi(\bm Z_1,...,\bm Z_4)\rangle}{\langle\Phi(\bm Z_1,...,\bm Z_4)|
 \Phi(\bm Z_1,...,\bm Z_4)\rangle},\\
 C_{i\rho j\sigma}&\equiv\frac{\partial^2\text{ln}\langle\Phi(\bm Z_1,...,\bm Z_4)|
 \Phi(\bm Z_1,...,\bm Z_4)\rangle}{\partial Z^*_{i\rho}\partial Z_{j\sigma}}.
\end{align}
By solving this EOM from an arbitrary initial wave function, a set of the vectors
$\bm Z_1(t),...,\bm Z_4(t)$ is obtained as a function of the real-time $t$, which
defines the basis wave function $\Phi(\bm Z_1(t),...,\bm Z_4(t))$ at each time.

When we solve the EOM, we add an external field $V_d$ to the Hamiltonian,
\begin{align}
 V_d&=v_d\sum_i f(|{\rm Re}\bm Z_i-\bm R_{c.m.}|),\\
 f(x)&=(x-a)^2\theta(x-a),\\
 \bm R_{c.m.} &=\frac{4}{13}\sum_{i=1}^3{\rm Re}\bm Z_i + \frac{1}{13}{\rm Re}\bm Z_4,
 \label{eq:wall}
\end{align}
where $\theta(x-a)$ is the step function with $a=10$ fm and  $v_d=1.5$ $\text{MeV/fm}^2$. This
external field reflects constituent particles at distance $a$ to prevent them escaping far away. 

Once we obtain a set of basis wave functions, we perform the generator coordinate method (GCM)
calculation by superposing them after the projection of the parity and angular momentum,
\begin{align}
 \Psi^{J\pi}_M=\int_0^{T_{\text{max}}}dt \sum^J_{K=-J}\hat{P}_{MK}^{J\pi}f_K(t)
 \Phi(\bm Z_1(t),...,\bm Z_4(t)),\label{eq:gcmwf}
\end{align}
where $\hat{P}^{J\pi}_{MK}$ is the parity and the angular momentum projection operator,
\begin{align}
 \hat{P}^{J\pi}_{MK}=\frac{2J+1}{8\pi^2}\int
 d\Omega\mathcal{D}_{MK}^{J*}(\Omega)\hat{R}(\Omega)\frac{1+\pi \hat{P}_x}{2}.
\end{align}
In the practical calculation, the integral in Eq.~(\ref{eq:gcmwf}) is discretized as,
\begin{align}
 \Psi^{J\pi}_M = \sum_{iK}\hat{P}^{J\pi}_{MK}f_{iK}\Phi_i, \label{eq:gcmwf2}
\end{align}
where $\Phi_i$ is an abbreviation for $\Phi(\bm Z_1(t),...,\bm Z_4(t))$. The amplitude $f_{iK}$ and
eigenenergy are determined by solving the Hill-Wheeler equation \cite{Hill1953,Griffin1957}.

\section{results and discussion}
\subsection{time evolution of the $\bm{3\alpha}$+$\bm n$ system} The numerical calculations were
performed according to the following procedure. First, using pure imaginary-time $\tau=it$ in
Eq.~(\ref{eq:eom}), we calculate the minimum intrinsic energy, that is found to be $-83.1$
MeV. Then, we generate the wave functions with the intrinsic excitation energy $E^*_{\rm int}$ using
the same equation. We have tested several excitation energies and used $E^*_{\rm int}=30$ MeV in
this work as it gives the best convergence of the GCM calculation. Using these wave functions
as the initial condition at $t=0$, we calculate the time evolution of the $3\alpha$+$n$ system.
The total propagation time was set to 10,000 fm/c, and the wave functions are recorded at every 33
fm/c. Consequently, an ensemble of the 300 wave functions is generated. By using different inital
wave functions at $t=0$, we generated two ensembles which we call set 1 and 2.
\begin{figure}[htb!]
 \includegraphics[width=0.7\hsize]{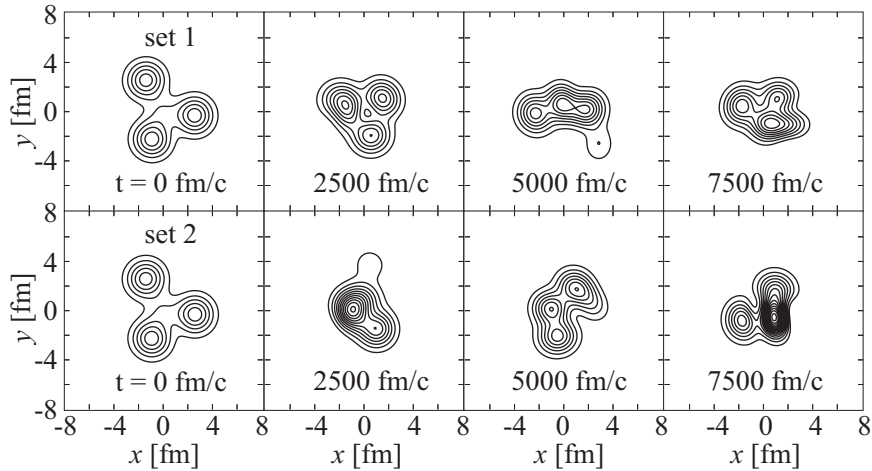}
 \caption{The snapshots of the intrinsic density distributions obtained by the
 real-time evolution. The top (bottom) panels show the wave functions from the ensemble
 set 1 (set 2). }  \label{fig:contour}
\end{figure}

\begin{figure}[hbt!]
 \includegraphics[width=0.6\hsize]{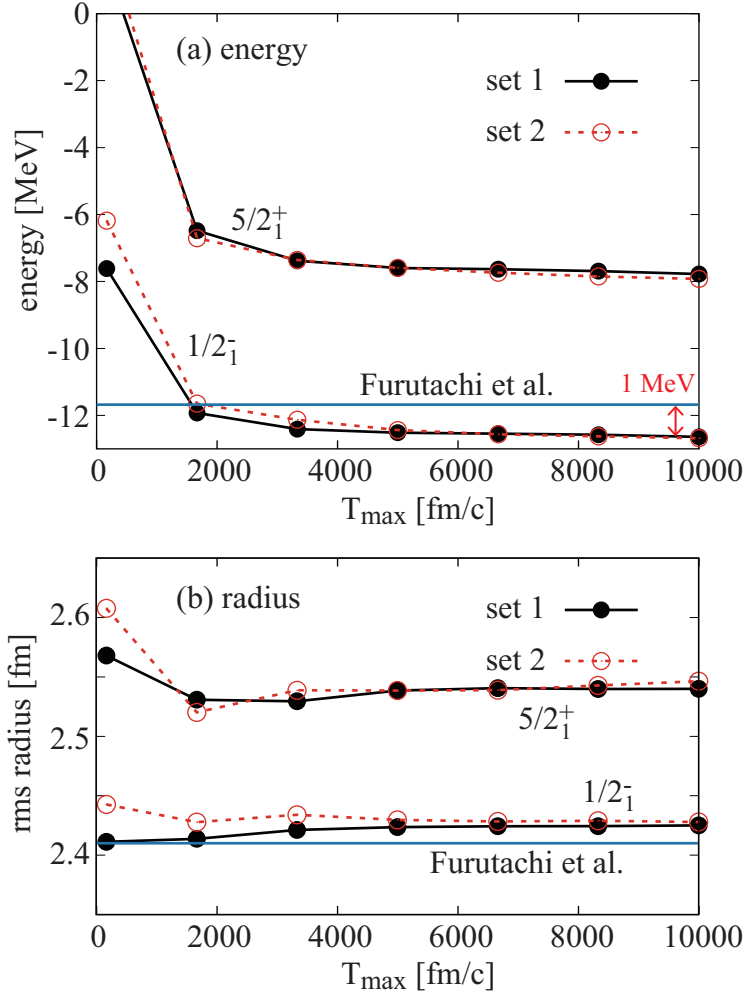}
 \caption{The energies and radii of the $1/2_1^-$ and $5/2_1^+$ states obtained from set
 1 and 2 as a function of the  total propagation time $T_\text{max}$. The strength 
 of the spin-orbit potential $u_{ls}=2000$ MeV was adopted. The result for the
 $1/2_1^-$ state obtained in Ref.~\cite{Furutachi2011} are denoted by blue lines.}
 \label{fig:conv}
\end{figure}

\begin{figure*}[hbt!]
 \includegraphics[width=\hsize]{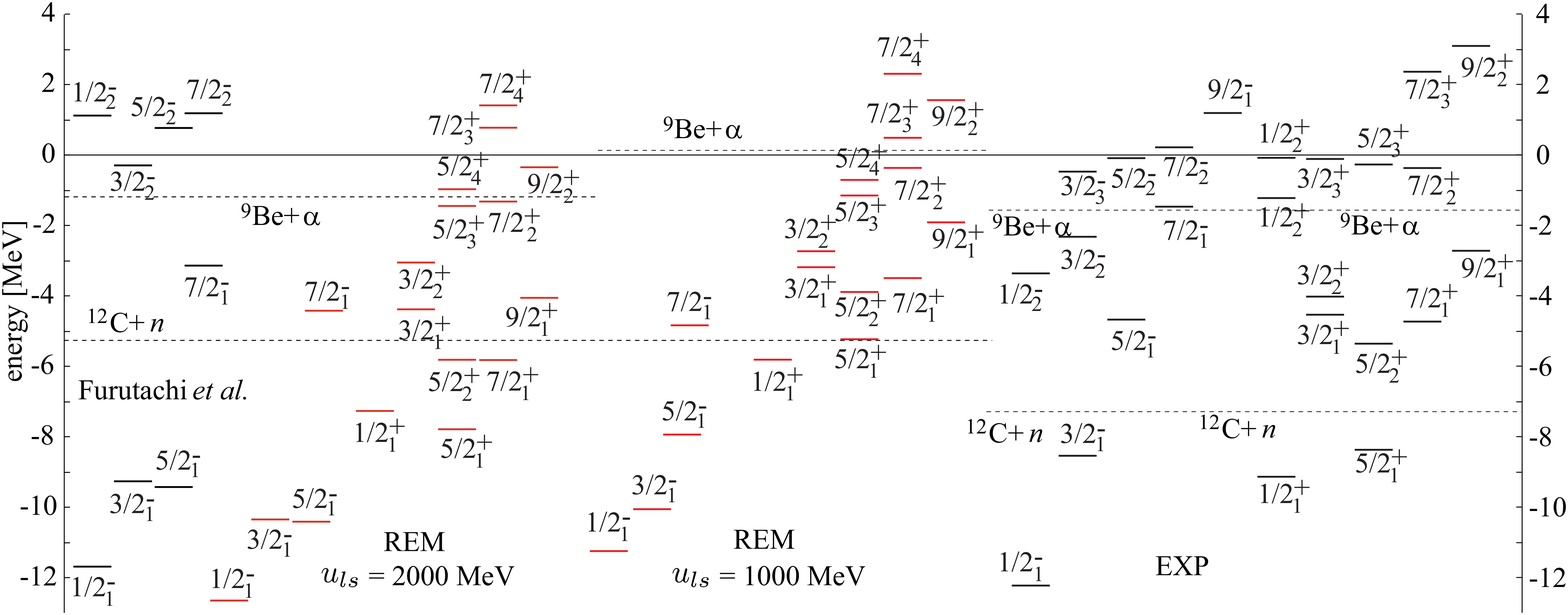}
 \caption{The energy spectrum of $^{13}$C calculated by using the strength of the spin-orbit
 potential $u_{ls}$=2000 and 1000 MeV. The energy is measured relative to the 3$\alpha+n$
 threshold. The spectrum is compared with that obtained by Furutachi et~al.~\cite{Furutachi2011}
 using  the same Hamiltonian with $u_{ls}$=2000 MeV.
 Experimental data are taken from Ref. \cite{Ajzenberg-Selove1991}. }
 \label{fig:level}
\end{figure*}

Several snapshots of the wave functions from these ensembles are shown in
Fig.~\ref{fig:contour}. Note that the wave functions of set 1 and 2 at $t=0$ fm/c have different
momenta of clusters, although they have almost the same spatial distributions. Consequently, the set
1 and 2 show the different results of the time evolution. We also note that various nuclear shapes
with different cluster configurations naturally emerge from the EOM. In some cases 3$\alpha$
particles are close to each other and the valence neutron is apart from them. In other 
cases, 2$\alpha$ particles and the valence neutron are close to each other, and an $\alpha$ particle
is apart from others describing $^9\text{Be}^*+\alpha$ like configurations. In this manner, the
ensembles of the basis wave functions were prepared without any assumption of the spatial symmetry. 

\subsection{the calculated full spectrum}
The generated wave functions are superposed to diagonalize the Hamiltonian. To confirm the
convergence of the calculation, Fig.~\ref{fig:conv} shows the  energies and radii of the
$1/2_1^-$ and $5/2_1^+$ states, which are the lowest negative- and positive-parity stats, as
functions of the propagation time $T_\text{max}$. The energy and radius of the ground state
($1/2^-_1$ state) show fast convergence and both sets reach almost the identical values. Thus, the
obtained GCM wave functions are converged well independent of the initial wave functions. The figure
also shows that REM yields approximately 1 MeV deeper binding energy of the $1/2^-_1$ state 
than the previous study by Furutachi et al.~\cite{Furutachi2011} who used the same Hamiltonian. This
clearly shows that REM can describe the $3\alpha$+$n$ system more accurately. It is interesting  
to note that REM gives the larger radius of the ground state despite the deeper binding energy. This
means that REM yields more stretched and long-ranged  wave function. It is also noted that
good convergence of the $5/2^+_1$ state was also achieved by using the same ensembles. 

The left half of Fig.~\ref{fig:level} compares the full spectrum obtained by REM and the
negative-parity states calculated by Furutachi et al.~\cite{Furutachi2011}. Because two calculations 
use the same Hamiltonian, deeper binding energy means a better description of the bound states below
the neutron threshold. Obviously, the present calculation gives deeper energies to all
the negative-parity states below the threshold ($1/2_1^-$, $3/2_1^-$ and $5/2_1^-$). It
also gives deeper binding energy to the $7/2^-_1$ state located just above the threshold, to which
the bound-state approximation may be validated. Thus, REM offers a better description of the bound
states than ordinary GCM calculations.   

However, the situation is different for the negative-parity resonances above the neutron threshold to
which variational principle is not applicable  and the bound-state approximation does not guarantee
the  energy convergence. In fact, two calculations disagree in the highly excited negative-parity
states. It is noted that the model space of REM is much larger than that of the GCM by Furutachi et
al.~\cite{Furutachi2011}. As a result, we found that most of the negative-parity resonances are coupled with
the non-resonant continuum which makes it difficult for us to identify resonant solutions from many
other non-resonant solutions. Therefore, we have not shown the negative-parity states above the
neutron threshold in Fig.~\ref{fig:level}. On the  contrary, although we cannot tell the reason
clearly, we found that the coupling is not strong in the positive-parity states, and stable
solutions are obtained which are plotted as resonances in the figure.   

The spectrum obtained by the spin-orbit strength $u_{ls}=2000$ MeV does not reproduce the order of
the ground band spectrum. It underestimates the excitation energy of the $5/2^-_1$ state and the
spectrum deviates from the observed rotational pattern.  This may affect the assignment of the
rotational bands and the discussion of the intrinsic shape. Therefore, we performed an additional
calculation using weaker spin-orbit strength $u_{ls}=1000$ MeV to check the interaction dependence
of the spectrum. As seen in Fig.~\ref{fig:level}, the weaker spin-orbit strength yields the correct
order of the ground band  member states ($1/2^-_1$, $3/2^-_1$, $5/2^-_1$ and $7/2^-_1$ states),
although it still overestimates the moment-of-inertia of the ground band. The side effect of the
weaker spin-orbit interaction is the overestimation of the excitation energies of the
positive-parity states. This may be due to the overestimation of the $^{9}{\rm Be}$+$\alpha$
threshold energy. If we measure them relative to the $^{9}{\rm Be}$+$\alpha$ threshold, the
excitation energies of many positive-parity states get closer to the observed values. This implies
that many positive parity-states have  $^{9}{\rm Be}$+$\alpha$ structure~\cite{Milin2002,Freer2011}.

\subsection{band assignment and shape of intrinsic states}

\begin{figure*}[tb!]
 \includegraphics[width=0.9\hsize]{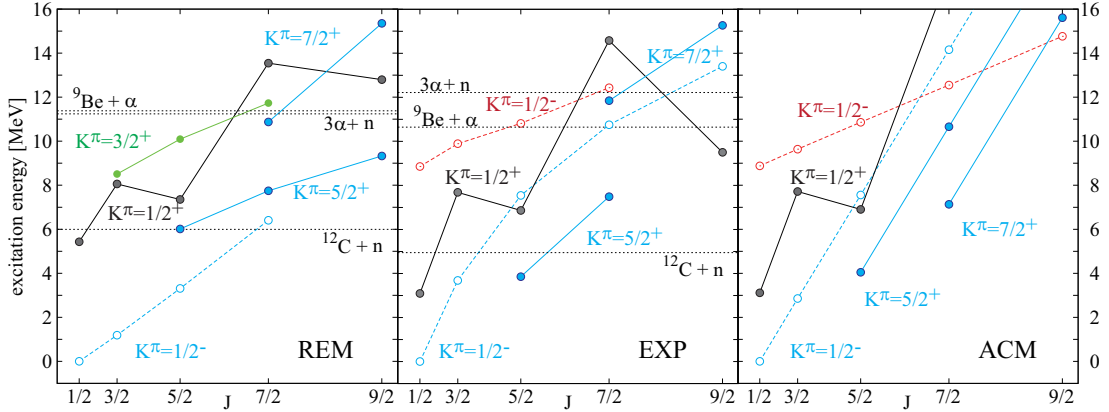}
 \caption{The band-assignment based on the calculated $E2$ transition strengths compared with that
 from the algebraic cluster model (ACM)~\cite{Bijker2019} and the experimental assignment which was
 also tentatively proposed in Ref.~\cite{Bijker2019}. The filled (open) symbols show the positive-parity
 (negative-parity) states.} \label{fig:band} 
\end{figure*}

\begin{table}[thb]
\caption{The calculated intra- and inter-band $E2$ transition probabilities in the unit of $e^2\rm
 fm^4$. The transitions larger than the Weisskopf estimate ($1 {\rm W.U.}=1.8\ e^2{\rm fm^4}$) are
 shown. The numbers in the parenthesis are the experimental values. }\label{tab:be2}  
\begin{ruledtabular} 
\begin{tabular}{llll}
 band $K^{\pi}_i \rightarrow K^{\pi}_f$ & $J_i$ & $J_f$ & $B(E2;J_i\rightarrow J_f)$ \\\hline
 $1/2^- \rightarrow 1/2^-$ & $1/2^-_1$ & $3/2^-_1$  & 17.4 (12.7)\\
                           &           & $5/2^-_1$  & 17.1 (16.9)\\
                           & $3/2^-_1$ & $5/2^-_1$  & 2.4  \\
                           &           & $7/2^-_1$  & 17.8 \\
                           & $5/2^-_1$ & $7/2^-_1$  & 2.0  \\
 $5/2^+ \rightarrow 5/2^+$ & $5/2^+_1$ & $7/2^+_1$  & 13.8 \rule{0pt}{10pt}\\
                           &           & $9/2^+_1$  & 10.9 \\
                           & $7/2^+_1$ & $9/2^+_1$  & 12.0 \\
 $7/2^+ \rightarrow 7/2^+$ & $7/2^+_2$ & $9/2^+_3$  & 12.9 \rule{0pt}{10pt}\\
 $1/2^+ \rightarrow 1/2^+$ & $1/2^+_1$ & $3/2^+_1$  & 16.7 \rule{0pt}{10pt}\\
                           &           & $5/2^+_2$  & 20.1 \\
                           & $3/2^+_1$ & $5/2^+_2$  & 5.0  \\
                           &           & $7/2^+_4$  & 7.6 \\
                           & $5/2^+_2$ & $9/2^+_2$  & 9.9 \\
 $3/2^+ \rightarrow 3/2^+$ & $3/2^+_2$ & $5/2^+_3$  & 10.0 \rule{0pt}{10pt}\\
                           &           & $7/2^+_3$  & 8.7 \\
                           & $5/2^+_3$ & $7/2^+_3$  & 9.8 \\\hline
 $5/2^+ \rightarrow 7/2^+$ & $7/2^+_1$ & $7/2^+_2$  & 4.3 \rule{0pt}{10pt}\\
 $1/2^+ \rightarrow 5/2^+$ & $1/2^+_1$ & $5/2^+_1$  & 6.7 (9.0)\rule{0pt}{10pt}\\
                           & $5/2^+_2$ & $5/2^+_1$  & 3.6 \\
                           &           & $7/2^+_1$  & 4.3 \\
                           &           & $9/2^+_1$  & 3.3 \\
                           & $9/2^+_2$ & $7/2^+_1$  & 2.2 \\
 $1/2^+ \rightarrow 3/2^+$ & $3/2^+_1$ & $7/2^+_3$  & 2.3 \rule{0pt}{10pt}\\
                           & $7/2^+_4$ & $5/2^+_3$  & 3.0 \\
 $3/2^+ \rightarrow 5/2^+$ & $3/2^+_2$ & $5/2^+_1$  & 2.7 \rule{0pt}{10pt}
 \end{tabular}
\end{ruledtabular}
\end{table}

Figure~\ref{fig:band} presents the band assignment determined from the calculated $E2$ transition
strengths listed in Tab.~\ref{tab:be2} and compares it with those from the experiment and the ACM
calculation. The band assignment of the REM results is unambiguous as the intra-band $E2$
transitions are clearly stronger than the inter-band transitions.

The $K^\pi=1/2^-$ band is built on the $1/2^-_1$ ground state. The intra-band $E2$ transition
strengths are reasonably described and comparable with the experimental data for the
$1/2^-_1\rightarrow 3/2^-_1$ and $1/2^-_1\rightarrow 5/2^-_1$ transitions. Experimentally, the
ground band terminates at the $9/2^-_{1}$ state, but we could not identify the corresponding state
in our calculation. This may be  due to the high excitation energy of this state which causes the
strong coupling with the continuum and makes it difficult to separate this state within the
bound-state approximation. 

For the positive-parity states, we have assigned four rotational bands; $K^\pi=5/2^+$, $7/2^+$,
$1/2^+$ and $3/2^+$ which are built on the $5/2^+_1$, $7/2^+_2$, $1/2^+_1$ and $3/2^+_2$ states,
respectively. Experimentally, the $E2$ transition strength for the $1/2^+_1 \rightarrow 5/2^+_1$
transition has already been measured (9.0 $e^2{\rm fm^4}$)~\cite{Ajzenberg-Selove1991} and our
calculation gives comparable value (6.7 $e^2{\rm fm^4}$). However, no other $B(E2)$ data is
available, and the positive-parity band assignment has not been firmly established by the
experiments. 

In Ref.~\cite{Bijker2019}, based on ACM which assumes the $3\alpha+n$ cluster structure with
triangular symmetry, the authors proposed a band assignment (Fig.~\ref{fig:band} right panel). They  
proposed the $K^\pi=1/2^-$, $5/2^+$ and $7/2^+$ bands which share the same intrinsic structure, and
the $K^\pi=1/2^+$ and $1/2^-$ bands with different structure. They also tentatively classified the
observed states into the rotational bands as shown in the middle panel of Fig.~\ref{fig:band}. In
addition to these four bands, they also pointed out the possible existence of a pair of the
$K^\pi=3/2^\pm$ band approximately at $E_x=10$ MeV. Interestingly, the global structure of the four
bands; $K^\pi=1/2^-$, $5/2^+$, $5/2^+$ and $1/2^+$ qualitatively agrees with the REM results,
although there exist several differences, for example, the order of the bands are different and
several bands are missing. It is also noted that REM shows quantitatively better agreement with
the experiment. 

Since the REM calculation does not assume any spatial symmetry, it is interesting to investigate if
there exists the triangular symmetry behind these rotational spectra. In general, the wave function
of REM is a superposition of many basis wave functions with different configurations, and hence, we
need some measure to evaluate its intrinsic structure. For this purpose, we introduce the overlap
between the REM wave function and the basis wave functions defined as, 
\begin{align}
 O_i = \sum_{KK'} \braket{\Psi^{J\pi}_M|P^{J\pi}_{MK}\Phi_i}B^{-1}_{KK'}
 \braket{P^{J\pi}_{MK'}\Phi_i|\Psi^{J\pi}_M},\label{eq:ovlp}
\end{align}
where $B^{-1}$ is the inverse matrix of $B$ which is the overlap of projected basis wave functions, 
\begin{align}
 B_{KK'} = \braket{P^{J\pi}_{MK}\Phi_i|P^{J\pi}_{MK'}\Phi_i}.
\end{align}
Note that the REM wave function $\Psi_{M}^{J\pi}$ is a superposition of $\Phi_i$
[Eq. (\ref{eq:gcmwf2})]. Therefore, if the overlap $O_i$ is large,  $\Psi_i$ can be approximated by
a single basis wave function $\Phi_i$.  
\begin{table}[thb]
 \caption{The calculated overlaps for each state which is defined by Eq. (\ref{eq:ovlp}). The columns
 denoted by $O(1/2^-)$ and $O(1/2^+)$ show the overlap between REM wave function and the basis wave
 function which is most dominant in the $1/2_1^-$ and $1/2_1^+$ states, 
 respectively.}\label{tab:ovlp}
 \begin{ruledtabular}
   \begin{tabular}{llllll}
    \multicolumn{3}{c}{$K^\pi=1/2^-$}&\multicolumn{3}{c}{$K=1/2^+$}\\
    $J^\pi$  & $O(1/2^-_1)$ & $O(1/2^+_1)$ &$J^\pi$  & $O(1/2^-_1)$ & $O(1/2^+_1)$ \\
    \cline{1-3}\cline{4-6}
    $1/2^-_1$ & 0.83 & 0.12 & $1/2^+_1$ & 0.14 & 0.58 \rule{0pt}{10pt}\\
    $3/2^-_1$ & 0.83 & 0.16 & $3/2^+_1$ & 0.18 & 0.56 \\
    $5/2^-_1$ & 0.73 & 0.06 & $5/2^+_2$ & 0.25 & 0.56 \\
    $7/2^-_1$ & 0.76 & 0.13 & $7/2^+_4$ & 0.35 & 0.25 \\
              &      &      & $9/2^+_2$ & 0.35 & 0.45 \\\hline
    \multicolumn{3}{c}{$K^\pi=5/2^+$}&\multicolumn{3}{c}{$K=7/2^+$}\rule{0pt}{10pt}\\
    $J^\pi$  & $O(1/2^-_1)$ & $O(1/2^+_1)$ &$J^\pi$  & $O(1/2^-_1)$ & $O(1/2^+_1)$ \\
    \cline{1-3}\cline{4-6}
    $5/2^+_1$ & 0.50 & 0.45 & $7/2^+_2$  &  0.74  & 0.15 \rule{0pt}{10pt}\\
    $7/2^+_1$ & 0.54 & 0.42 & $9/2^+_3$  &  0.55  & 0.19 \\
    $9/2^+_1$ & 0.58 & 0.43 &            &        &      \\\hline
    \multicolumn{3}{c}{$K^\pi=3/2^+$}\rule{0pt}{10pt}\\
    $J^\pi$  & $O(1/2^-_1)$ & $O(1/2^+_1)$ \\\cline{1-3}
    $3/2^+_2$ & 0.45 & 0.40 \rule{0pt}{10pt}\\
    $5/2^+_3$ & 0.46 & 0.26 \\
    $7/2^+_3$ & 0.35 & 0.28 
   \end{tabular}
 \end{ruledtabular}
\end{table}
\begin{figure}[tb!]
 \includegraphics[width=0.5\hsize]{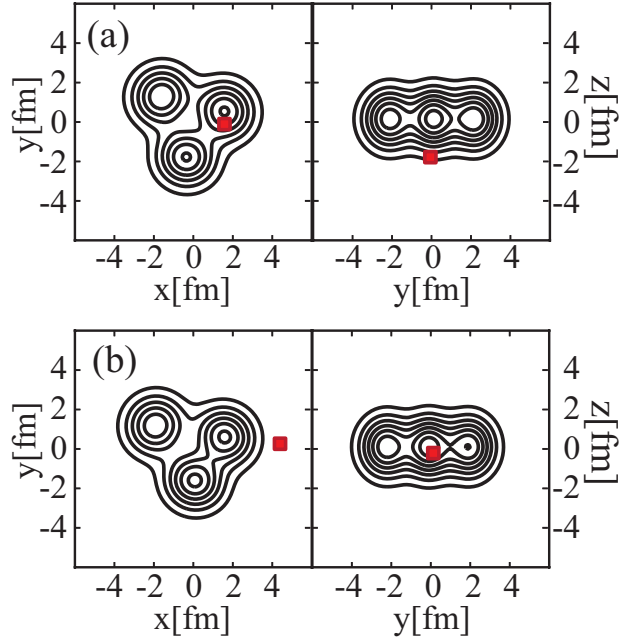}
 \caption{Panel (a): the density distribution of the basis wave functions which have the maximum
 overlap with the $1/2^-_1$ state. Panel (b): Same as panel (a) but for the $1/2^+_1$
 state. Contours show the density of $3\alpha$ particles and red boxes show the position of the
 wave packet centroid for the valence neutron. }
 \label{fig:ovdens}   
\end{figure}

The calculated overlaps are summarized in Tab.~\ref{tab:ovlp}. The ground state has the maximum
overlap, which is as large as 0.83, with the basis wave function shown in Fig.~\ref{fig:ovdens}
(a). Note that the density distribution clearly shows the triangular configuration of $3\alpha$ 
particles with a valence neutron where the lengths of the triangle are 3.31, 3.30 and 3.02
fm. Furthermore, we found that all the member states of the ground band have large overlaps no less
than 0.70 with the same basis wave function. Therefore, we consider that the ground band is
reasonably interpreted as the rotational band having a common intrinsic structure with a triangular
symmetry as asserted by Bijker et al.~\cite{Bijker2019}. They also argued that the $K^\pi=5/2^+$ and
$7/2^+$ bands have the same intrinsic structure and are classified as the ``ground band''. Indeed,
we found that these bands have non-small overlap with the same basis wave function shown in
Fig.~\ref{fig:ovdens} (a). However, our results show a deviation from a rigid shape. The
magnitudes of the overlaps between these bands and the basis wave function shown in
Fig.~\ref{fig:ovdens} (a) are reduced less than 0.60 except for the $7/2^+_2$ state. Furthermore, 
these bands have non-small overlaps with other configurations. For example, the $K^\pi=5/2^+$ band
has large overlap with the dominant basis wave function of the $1/2^+_1$ state, which is discussed
below. Thus, the $K^\pi=5/2^+$ and  $7/2^+$ bands look similar to the $K^\pi=1/2^-$ band, but the
deviation from the rigid shape is  not small.

In Ref.~\cite{Bijker2019}, the $K^\pi=1/2^+$ band was assigned as a rotational band which also has a 
triangular arrangement of $3\alpha$ particles but has the valence neutron in a different
single-particle orbit. In the present calculation, we also found that the band-head state  
($1/2^+_1$ state) has the maximum overlap with a different basis wave function whose density
distribution is shown in Fig.~\ref{fig:ovdens} (b), but has small overlap with the dominant
configuration of the ground band [Fig.~\ref{fig:ovdens} (a)]. Apparently, the position of the wave 
packets of the valence neutron is different from that of the $1/2^+_1$ state, and $\alpha$ particles
deviate from equilateral triangular arrangement as the lengths of the triangle are 3.55, 3.51 and
2.67 fm. This confirms that the $K^\pi=1/2^+$ band has a different intrinsic structure. However, we
again note that the magnitude of the maximum overlap is not as large as that of the ground band, and
the member states of this band show the increasing mixture of other contributions as the
excitation energy and angular momentum increase. Finally, we also found the 
strongest admixture of the various configurations in the $K^\pi=3/2^+$ band which is a candidate of
the band proposed in Refs. \cite{Milin2002,Bijker2019}. This may be due to the highest high
excitation energy of this band.

In short, the REM calculation confirmed that the ground band can be interpreted as a
rigid-body rotational band which manifests the triangular symmetry. It also shows that ACM
looks explaining the general trend of the excited bands. However, we found that all the
excited bands have non-small admixture with other configurations and deviate from the
rigid-body interpretation. One of the signature of this mixing is the non-small $E2$ transitions between the bands with different intrinsic structures. Therefore, the experimental data
for these transitions will provide us an important insight into the cluster structure of $^{13}{\rm
C}$.

\section{summary}
In summary, we have investigated the structure of the $3\alpha+n$ system by extending the REM
framework. As a benchmark calculation for the 3$\alpha+n$ system, REM well reproduced the ground and
excited energies where we followed the same Hamiltonian of the previous study as a comparison.  It
was also demonstrated that REM accurately describes the wave functions which yields to the deeper 
binding energies.   

We have also discussed the rotational band assignment and investigated if they manifest the
triangular symmetry. The proposed band assignment qualitatively explains the observed data, although 
the order of several bands disagrees and the $K^-=1/2^-$ band is missing in the present result. From
the analysis of the overlap with the basis wave functions, it was found that the ground band can be
regarded as a rigid-body rotational band which manifests the triangular symmetry. We also have seen
that the $D'_{3h}$ symmetry approximately explains the general nature of the excited bands. However,
all the excited  bands have non-small admixture with other 
configurations without symmetry and deviate from the rigid-body interpretation, because of their
high excitation energies and angular momenta. The non-small $E2$ transitions between different
bands are a signature of the configuration mixing, and we expect that the experimental
data for these transitions will provide us an important information about the underlying
symmetry behind the observed spectrum.

\begin{acknowledgements}
 The authors acknowledge the fruitful discussions with Dr. Funaki and  Dr. Kawabata. This work was
 supported by JSPS KAKENHI Grant Nos. 19K03859, the collaborative research programs 2020 at the
 Hokkaido University information initiative center, and by the COREnet program at the RCNP,
 Osaka University.
\end{acknowledgements}

\bibliography{main}
\end{document}